\documentclass[11pt]{article}

\begin{document}
\thispagestyle{empty}

\title{COMPLEX ANGULAR MOMENTA\\
AND THE PROBLEM OF EXOTIC STATES}

\author{ Ya.I. AZIMOV\\
\\
Petersburg Nuclear Physics Institute,\\
Gatchina, 188300, Russia}
\maketitle


\begin{abstract}
After some brief personal recollections about V.N.Gribov, I demonstrate
that his results and ideas on complex angular momenta may be applied
in unfamiliar directions. As an example, it is shown, that any strong
interaction amplitude, satisfying dispersion relation (in momentum
transfer), has infinite number of energy-plane poles, both for exotic
and non-exotic quantum numbers. This result ensures the necessary
condition for existence of exotic hadrons. However, without more
detailed knowledge of dynamics one cannot secure sufficient conditions
for the exotics existence.
\end{abstract}

\section{Introduction}

I was lucky to work for 20 years in the close contact with V.N.~Gribov,
in the Ioffe Physico-Technical Institute and in the Leningrad (now
Petersburg) Nuclear Physics Institute. It seems that I was the youngest
of his colleagues who called him by his Russian nickname, Volodya. When
he changed his home institution, to the Landau Theoretical Physics
Institute, our contacts still continued, though became irregular. Later,
after his death in 1997, I looked through his list of publications and,
quite unexpectedly, was disappointed. For me, both the list, and
publications themselves, could hardly reflect his bright personality.

V.N.~Gribov was a rare person, with whom one could discuss
practically any problem of physics (and not only physics).
Discussions with him were always very intensive and passionate.
Sometimes he was not right after all, but even then he suggested
such non-trivial arguments for his position, which were difficult
to reject. Similar discussions always emerged also at our
theoretical seminars, even if the presented work had been earlier
discussed with Gribov personally. I think that was why many
physicists disliked (feared?) to talk at our seminars. However,
being very difficult, discussions with Gribov were always very
interesting and very useful. They gave to the author a better
understanding of own work. Moreover, they provided a beautiful
training how to present results of the work.

References to Gribov's works are rather frequent today in the
literature. But they mostly refer to QCD problems only - Gribov
copies, Gribov horizon, and, of course, DGLAP equations for the
parton evolution (here G means just Gribov). Very intensive Gribov's
activity on Regge theory seems to be nearly forgotten now. Meanwhile,
his numerous papers in this direction may still generate ideas and
arguments on various problems, even if the problems look totally
unrelated to the Reggeology. In the present note I demonstrate this
for the problem of exotic states, \textit{i.e.}, hadrons with such
quantum numbers that cannot be composed of 3 quarks (for baryons)
or one quark-antiquark pair (for mesons). The corresponding result
has been published earlier~\cite{aaswg}. Its derivation and meaning
will be discussed here in more detail.

\section{
Dispersion relations and energy-plane poles}

As is well-known, amplitude of any 2-particle-into-2-particle
process depends on two kinematical variables. They may be taken
to be $s=W^2$ and $z=\cos\theta$, with $W$ and $\theta$ being
the c.m. energy and scattering angle, respectively. It is convenient
in many cases, instead of $z$, to use the invariant quantity,
squared c.m. momentum transfer. There are two such variables:
$t$, linear in $z$, and $u$, linear in $-z$ (of course, the sum
$t+u$ is independent of $z$).

As function of $z$, the invariant amplitude $A(s,z)$ can be
decomposed into partial-wave amplitudes $f_j(s)$, that correspond
to the process at a definite value of the total angular
momentum $j$, the same in initial and final states. Physical
values of $j$ are (half)integer. Of course, one could define
$f_j(s)$ as an analytical function for arbitrary values of $j$,
real or even complex, such that coincides with physical amplitudes
at physical values of $j$. Such procedure is always possible, but
is generally ambiguous (\textit{e.g.}, when continuing from positive
integer points one could add any term proportional to
$1/\Gamma(1-j)\,$).  However, under some conditions, the partial-wave
amplitudes $f_j(s)$ admit an unambiguous analytical continuation to
non-physical values of $j$.

In what follows, we use three assumptions: 1)~possibility of the
unambiguous analytical continuation in the angular momentum $j$;
2)~absence of massless physical hadrons; 3)~unitarity condition.
Let us consider their meaning.

For relativistic amplitudes, the possibility of unambiguous
analytical continuation in $j$ is usually connected with
dispersion relations in $z$ (see, \textit{e.g.}, the
monograph~\cite{col}). Generally, dispersion relations have not
been deduced mathematically from axioms of the Quantum Field
Theory. They have not been proved also in terms of the more
specific Quantum Chromodynamics (QCD), which is believed now to
provide the physical basis for strong interactions. However, not
all details of QCD have been understood up to now. Even relation
between quarks and gluons, on one side, and mesons and/or baryons,
on the other, cannot be described today in a model-independent
way. That is why various dispersion relations (or their
analogs/generalizations) are widely applied to phenomenological
treatment of strong interactions. The dispersion relations are
used as input in modern Partial-Wave Analyses (PWA; see,
\textit{e.g.}, the latest pion-nucleon PWA~\cite{piN}). They are
operative to pick out such strong interaction parameters as
meson-baryon coupling constants, and/or the pion-nucleon
$\sigma$-term. Analytical properties, corresponding to dispersion
relations, provide basis for various sum rules.  Analogous
analytical properties of inelastic amplitudes are assumed when
extracting the meson-meson scattering amplitude from data on
processes like $\pi\to 2\pi$. Up to now, phenomenological
applications of dispersion relations have not revealed any
inconsistencies, neither theoretical nor experimental, and we
still may consider them to be true.

Dispersion relations lead to the integral Gribov-Froissart
formula~\cite{GF,col} for the continued partial-wave amplitudes,
at least at large positive Re$\,j$. Note, however, that the
dispersion relations provide a sufficient, but not necessary
condition for the analytical continuation of $f_j(s)$ in $j$. The
continuation survives under much weaker conditions on
singularities of the invariant amplitudes in $t$ and/or $u$, and
the continued partial-wave amplitudes still conserve essentially
the same Gribov-Froissart form.

Now, consider the 2nd condition. Let us begin, for simplicity,
with the amplitude for elastic scattering of two spinless particles,
where the total angular momentum $j$ coincides with the orbital
momentum $l$. If there are no massless hadrons, the physical
partial-wave amplitudes, being described by the Gribov-Froissart
formula, reveal the familiar threshold behavior
\begin{equation} f_l(s)\sim k^{2l}~~~{\rm at}~~~k\to0\,,
\end{equation} where $k$ is the c.m. momentum. The same threshold
behavior is true also for the continued amplitudes $f_l(s)$ with
any, real or complex, angular momenta, at least till the
Gribov-Froissart formula is applicable. This result has very
simple physical meaning. Non-massless exchange, in the
non-relativistic limit, induces the Yukawa potential of the form
$\exp(-\mu r)/r$, with non-vanishing $\mu$ and the limited radius
of action $R\sim1/\mu$. As is well-known, every quantum mechanical
potential with final radius $R$ generates amplitudes with the
threshold behavior $\sim (kR)^{2l}$. In contrast, a massless
exchange, in the same limit, would correspond to the Coulomb
potential, with infinite radius and more complicated threshold
behavior.

Considering unitarity condition, we can again begin with the
simple case of purely elastic interaction near the threshold. Then
the unitarity relation, continued to any real $l$, conserves the
same form as for physical values of $l$: \begin{equation}
f_l(s)-f_l^*(s)=2ik\,f_l(s)\,f_l^*(s) \,. \end{equation}
Evidently, the threshold behavior (1) at real $l$ may be
consistent with the unitarity relation (2) only if $l\geq-1/2$.
This means that near the threshold (\textit{i.e.}, at $s\to s_{\rm
th},\,k^2\to0$) the Gribov-Froissart formula cannot stay unchanged
(with its integral convergent) at  Re$\,l\leq-1/2$.

The analysis, made by Gribov and Pomeranchuk~\cite{GP}, shows that
at $k^2\to+0$ the analytically continued partial-wave amplitude
$f_l(s)$ should reveal unlimited number of poles in $l$ at
Re$\,l>-1/2$. They are Regge poles (or reggeons), with
energy-dependent positions. Further, from more detailed
consideration of the small-$k^2$ region, Gribov and Pomeranchuk
demonstrated~\cite{GP} that the reggeons in the vicinity of
$l=-1/2$ have trajectories \begin{equation} \label{grp}
l_n(s)\approx -\frac12 + \frac{i\pi n}{\ln(R\sqrt{|k^2|})} +
O(\ln^{-2}(R\sqrt{|k^2|})) \,, \end{equation} with $R$ being the
effective interaction radius. The number $n$ takes any positive
and negative integer values, $n=\pm1,\pm2,...\,$. We see that
there are infinitely many reggeons accumulating to $l=-1/2$ at
$k^2\to0$.

The structure of this accumulation, at real or even complex energies,
may be investigated more explicitly in the non-relativistic quantum
mechanics with final range potential, \textit{e.g.}, with the Yukawa
potential~\cite{aas1,aas2}. Below the threshold (at $k^2<0$), the set
of accumulating poles consists of infinitely many pairs of reggeons in
the left half-plane Re$\,l\leq-1/2$. The reggeons of each pair have
the same $|n|$, and their trajectories are complex conjugate to each
other. At $k^2\to-0$, they tend to the limiting point -1/2.  Such
simple correlation becomes destroyed above the threshold (at $k^2>0$),
and some reggeons appear in the right half-plane Re$\,l\geq-1/2$.

Up to now we have neglected particle spins. Accounting for them
changes the situation, but only quantitatively. For the elastic
threshold of two particles with spins $\sigma_1$ and $\sigma_2$,
there appear several accumulation points in the $j$-plane, the
rightmost one is at \begin{equation} \label{jsp} j=-1/2+\sigma_1+
\sigma_2\,, \end{equation} instead of $j=-1/2$~\cite{az}. The
reason is simple: the accumulation points still correspond to
$l=-1/2$, but particle spins provide several possible $j$-values
for any fixed $l$-value, and \textit{vice versa}. Correspondingly,
structure of the accumulations in the $j$-plane, at thresholds of
two spinning particles, are still described by trajectories
(\ref{grp}), with the shifted limiting points (\ref{jsp}). The
case of multi-particle thresholds has never been really studied
(though hypothesized by Gribov and Pomeranchuk~\cite{GP}).

The above consideration was applied to a purely elastic two-particle
case, and the discussed reggeons were seen as poles for the
corresponding elastic partial-wave amplitudes. But at other
(higher) energies the unitarity relation connects the elastic
scattering with different, inelastic processes. After all, we arrive
at the conclusion, that partial-wave amplitudes of all processes, which
can be related by rescattering, should couple to the same reggeons.
In particular, they should contain those accumulating Regge poles.

Essential for our present purpose is the infinite number of the
accumulating Regge poles~\cite{GP}, which implies that the total
number of Regge poles is certainly infinite as well. Their
positions  depend on energy and are determined by some relation of
the form \begin{equation}  \label{f} F(s,\,j)=0\,. \end{equation}
Each of its solution corresponds to a pole of a partial-wave
amplitude and may be considered in two ways: either as a pole in
$j$ with the position depending on energy (on $s$), or as a pole
in energy (in $s$) with the position depending on the total
angular momentum $j$. This provides one-to-one correspondence
between reggeons and energy-plane poles.  Therefore, the infinite
number of reggeons corresponds to the infinite number of poles in
the energy plane. The same conclusion is true for the
non-relativistic case of a finite-range potential, in particular,
for the Yukawa potential. This allows to investigate more clearly
the pole structure of amplitudes.

At first sight, the infinite number of energy-plane poles for any
finite-range potential looks impossible, since the energy-plane poles
are related to bound states or resonances. As well known, the bound
or resonance states may exist, say, in the Yukawa potential, only if
it is attractive and sufficiently strong. However, one can ask here
an interesting question: whether every complex-energy pole should be
considered as possibly related with a resonance? The problem is that,
in terms of complex angular momenta, the Regge poles participating in
the Gribov-Pomeranchuk accumulations are clearly separated from physical
points, which are, for the $j$-plane, only integer (or half-integer)
non-negative points.  Thus, the accumulating poles cannot provide
physically meaningful (bound or resonance) states, at least near the
corresponding threshold. This could be true for non-threshold energies
as well.

Moreover, it might be that the total set of Regge poles is split into
two (or more?) different subsets: one related, say, with the
Gribov-Pomeranchuk accumulations, the other with the bound and/or
resonance states. Then there could be infinite number of poles of the
former type, while only few (if any) poles of the latter type. This
would be so, \textit{e.g.}, if Eq.(\ref{f}) could be factorized as
$$F_1(s,j)\,F_2(s,j)=0\,.$$ However, explicit expressions of
Ref.\cite{aas1} show non-factorizable structure of Eq.(\ref{f}).
Moreover, there is no basic difference between various Regge
poles~\cite{aas2}: all Regge trajectories appear to be different
branches of the same multi-valued analytical function. Formally, this
fact is related to non-trivial analytical properties of the Regge
trajectories as functions of the energy: their singularities (branch
points) come not only from physical thresholds, but also from
coincidence of two (or more) reggeons~\cite{aas2}, where those reggeons
can be interchanged.

When considering the poles as energy-plane poles, one should have in
mind that the energy plane generally has many Riemann sheets. An
energy-plane pole corresponds to a bound state or resonance only if
it is placed at the physical sheet or near the physical region. Poles
placed far from the physical region or at far Riemann sheets are
``not seen''.

The infinite number of reggeons for the Yukawa potential can be
``visualized''. Evidently, the limit $\mu\to0$ transforms the Yukawa
potential $\exp(-\mu r)/r$, with the finite radius of interaction
$\sim1/\mu$, into the Coulomb potential, with the infinite radius.
As traced in Ref.\cite{aas1}, this limit simultaneously transforms
the Yukawa potential reggeons, that realize the Gribov-Pomeranchuk
accumulation, into Coulomb reggeons, that realize orbital and radial
Coulomb excitations. Therefore, the infinite number of energy-plane
poles in the Yukawa potential becomes seen in the limit $\mu\to0$
as the infinite number of the Coulomb levels. The Gribov-Pomeranchuk
accumulation of reggeons near the threshold energy transforms in this
limit into the well-known accumulation of the Coulomb bound states to
the threshold (note that the double limiting transition $\mu\to0,\,
k\to0$ is not equivalent here to the similar, but reversed limit
$k\to0,\,\mu\to0$).

Returning from the non-relativistic quantum mechanics to relativistic
amplitudes, we emphasize that the above arguments have not assumed any
specific quantum numbers in the scattering channel. Therefore, their
conclusion on the infinite number of energy-plane poles should be
equally applicable (or non-applicable) to both bosonic and fermionic
hadron poles, having any flavor quantum numbers (exotic or non-exotic).

Thus, if we study $2\to2$ strong interaction amplitudes, we should
admit existence of (infinite number of) complex-energy poles with
any exotic quantum numbers, both mesonic and baryonic. Alternatively,
one could assume that analytical properties of strong interaction
amplitudes, having exotic quantum numbers (at least, for one of the
physical channels, $s$-, $t$-, or $u$-channel), are essentially
different from those of totally non-exotic amplitudes. Such assumption
looks very unnatural, and should be rejected. Thus, we obtain an
argument for existence of exotic states, absent in previous
publications.

Of course, existence of the energy-plane poles is only a necessary
condition for the physical existence of exotic hadrons. To guarantee
fulfillment of the sufficient condition, \textit{i.e.}, existence of
energy pole(s) with particular quantum numbers near the physical
region, one should have more detailed knowledge of dynamics.

It is interesting to note that the above consideration uses the
reggeons in a non-standard manner. Usually, to apply the complex
angular momenta approach for obtaining results in the $s$-channel
(the channel where the invariant $s$ has the meaning of the
squared c.m. energy), one begins from the crossed channel, where
$t$ and $s$ are, respectively, the squared energy and squared
momentum transfer (see Ref.~\cite{col}). Analytical continuation
of partial-wave amplitudes in this $t$-channel allows, after
returning into the $s$-channel, to study behavior of the invariant
amplitude (and cross section) at high energy $s$ at fixed value of
momentum transfer $t$. Now, to obtain the conclusion about
energy-plane poles in the $s$-channel, we use complex angular
momenta in the same $s$-channel.

To summarize, earlier results on complex angular momenta imply
that any 2-hadron interaction amplitude, under standard
assumptions, has infinite number of poles in the energy plane.
Those poles may have exotic, as well as non-exotic, quantum
numbers. However, the poles may be placed at far Riemann sheets of
the energy plane. Therefore, more detailed dynamical information
is necessary to guarantee real existence of exotic hadrons.

\section*{Acknowledgments}
In preparing this work I was strongly influenced by numerous past
discussions with V.N.~Gribov. I also thank R.A.~Arndt,
I.I.~Strakovsky, R.L.~Workman, and K.~Goeke, my coauthors in
paper~\cite{aaswg}, for more recent stimulating discussions. This
text is based on the work partly supported by the U.~S.~Department
of Energy Grant DE--FG02--99ER41110, by the Jefferson Laboratory,
by the Southeastern Universities Research Association under DOE
Contract DE--AC05--84ER40150, by the Russian-German Collaboration
(DFG, RFBR), by the COSY-Juelich-project, and by the Russian State
grant RSGSS-1124.2003.2.


\end{document}